\documentclass[graphics,floatfix, footinbib,tightenlines,nobibnotes, aps, prb, twocolumn]{revtex4-1}

\usepackage{amsmath,amssymb,wasysym}
\usepackage{graphicx}
\usepackage{dcolumn}
\usepackage{bm}
\usepackage{braket}
\usepackage{verbatim}
\usepackage{float}
\usepackage{bbold}
\usepackage{color}
\usepackage{xcolor}
\usepackage{relsize}
\usepackage{amsthm}
\usepackage{enumerate}
\usepackage{soul,xcolor}
\usepackage[T1]{fontenc}
\usepackage{adjustbox}
\usepackage[colorlinks=true ,urlcolor=blue,urlbordercolor={0 1 1}]{hyperref}

\newcommand{\be}{\begin{equation}}
\newcommand{\ba}{\begin{align}}
\newcommand{\ee}{\end{equation}}
\newcommand{\bea}{\begin{eqnarray}}
\newcommand{\eea}{\end{eqnarray}}
\newcommand{\beq}{\begin{equation}}
\newcommand{\eeq}{\end{equation}}
\newcommand{\beqn}{\begin{eqnarray}}
\newcommand{\eeqn}{\end{eqnarray}}

\newcommand*{\yhz}[1]{\textcolor{black}{#1}}

\usepackage{bm}

\usepackage{array}
\newcolumntype{L}[1]{>{\raggedright\arraybackslash}p{#1}}
\newcolumntype{C}[1]{>{\centering\arraybackslash}p{#1}}
\newcolumntype{R}[1]{>{\raggedleft\arraybackslash}p{#1}}
\usepackage{multirow}

\allowdisplaybreaks

\begin{document}

\title{Type II t-J model and shared super-exchange coupling from Hund's rule in superconducting La$_3$Ni$_2$O$_7$ }

\author{Hanbit  Oh}
\author{Ya-Hui Zhang}
\email{yzhan566@jhu.edu}
\affiliation{William H. Miller III Department of Physics and Astronomy, Johns Hopkins University, Baltimore, Maryland, 21218, USA}

\date{\today}

\begin{abstract}
Recently, an 80 K superconductor was discovered in La$_3$Ni$_2$O$_7$ under high pressure. 
Density function theory (DFT) calculations identify $d_{x^2-y^2}$, $d_{z^2}$ as the active orbitals on the bilayer square lattice 
with a $d^{8-x}$ configuration of Ni per site. Here $x$ is the hole doping level.
One naive expectation is to describe this system in terms of a two-orbital t-J model. 
However, we emphasize the importance of Hund's coupling $J_H$ and the $x=0$ limit should be viewed as a spin-one Mott insulator.
Especially, the significant Hund's coupling shares the inter-layer super-exchange $J_\perp$ of the $d_{z^2}$ orbital to the $d_{x^2-y^2}$ orbital, an effect that cannot be captured by conventional perturbation or mean-field approaches.
This study first explores the limit where the $d_{z^2}$ orbital is Mott localized, dealing with a one-orbital bilayer t-J model focused on the $d_{x^2-y^2}$ orbital. 
Notably, we find that strong inter-layer pairing survives up to $x=0.5$ hole doping driven by the transmitted $J_\perp$, which explains the existence of a high Tc superconductor in the experiment at this doping level.
Next, we uncover the more realistic situation where the $d_{z^2}$ orbital is slightly hole-doped and cannot be simply integrated out. 
We take the $J_H\rightarrow +\infty$ limit and propose a type II t-J model with four \textit{spin-half} singlon ($d^7$) states and three \textit{spin-one} doublon ($d^8$) states. 
Employing a parton mean-field approach, we recover similar results as in the one-orbital t-J model, but now with the effect of the $J_\perp$ automatically generated.
Our calculations demonstrate that the pairing strength decreases with the hole doping $x$ and $x=0.5$ is likely larger than the optimal doping. We propose future experiments to electron dope the system to further enhance $T_c$. 
 \end{abstract}

\maketitle
\textit{Introduction}:  Recently a superconductor with $T_c=80$K was found in La$_3$Ni$_2$O$_7$ under high pressure\cite{sun2023signatures}, following  previous discoveries of  superconductivity in nickelate Nd$_{1-x}$ Sr$_x$NiO$_2$ \cite{li2019superconductivity} and also in Nd$_6$Ni$_5$O$_{12}$\cite{pan2022superconductivity} at ambient pressure. The discovery has triggered many  experimental\cite{liu2023electronic,hou2023emergence} and theoretical\cite{luo2023bilayer,zhang2023electronic,yang2023possible,sakakibara2023possible,gu2023effective,shen2023effective,wu2023charge,christiansson2023correlated,liu2023s,hou2023emergence,liu2023electronic,cao2023flat} studies.
The average valence of Ni is in $d^{8-x}$ with \textcolor{black}{hole doping level,} $x=0.5$\cite{sun2023signatures}. Density functional theory (DFT) calculations identify a bilayer square lattice structure with active $d_{x^2-y^2}$ and $d_{z^2}$  orbitals, which we label as $d_1$ and $d_2$ in the following. The density (summed over spin) per site is estimated to be $n_1\approx 1-x=0.5$ and $n_2 \approx 1$, so that the $d_{z^2}$ orbital is close to Mott localization. Due to a large inter-layer hybridization of the $d_{z^2}$ orbital, we expect that it just forms a rung singlet when $n_2=1$.    The $d_{z^2}$ orbital has a small intra-layer hopping, thus we do not expect a strong superconductivity from it.  Then one may expect that superconductivity originates from the $d_{x^2-y^2}$ orbital. But the $d_{x^2-y^2}$ orbital is at hole doping level of $50\%$. According to the phase diagram of cuprates, it should be in the overdoped Fermi liquid phase.  A major goal of this paper is to identify the minimal model to describe the nickelate superconductor and also find a mechanism for the material to superconductor at such a large hole doping.

One important ingredient we identify is Hund's coupling $J_H$ between the $d_{z^2}$ and the $d_{x^2-y^2}$ orbital. Due to the $J_H$ coupling, the $x=0$ limit should be viewed as a spin-one Mott insulator formed by Ni$^{2+}$. The strong Hund's coupling $J_H$ aligns the spin of the two orbitals at each site, then the large inter-layer spin coupling $J_\perp$ of the $d_{z^2}$ orbital is shared to the $d_{x^2-y^2}$ orbital. Therefore, when $n_2=1$, we can ignore the Mott localized $d_{z^2}$ orbital (which is in a gapped rung-singlet phase) and phenomenologically consider a bilayer one-orbital t-J model for $d_{x^2-y^2}$ only.  The model has  a large inter-layer spin coupling $J_\perp$ but without inter-layer hopping $t_\perp$, a new situation not possible in the usual one-orbital Hubbard model. Through a slave-boson mean field calculation, we find that a large $J_\perp$ disfavors the familiar $d_{x^2-y^2}$ pairing at the $J_\perp=0$ limit and the system forms a strong s-wave superconductor with dominant inter-layer pairing. The pairing strength decreases with the hole doping level $x$. But with a sufficiently large $J_\perp$, the pairing survives at $x=0.5$, which explains the superconductor at this hole doping level in the experiment. We note that a previous work has discussed quantitative renormalization effects of the Hund's coupling in flattening the bands\cite{cao2023flat}, but the effect we identify here is qualitatively distinct and completely new. To our best knowledge the possibility  of strong inter-layer pairing for the $d_{x^2-y^2}$ orbital due to Hund's rule coupling to a rung-singlet phase of the $d_{z^2}$ orbital has not been discussed previously.

The above treatment of `integrating' out the $d_{z^2}$ orbital is not very rigorous. Also, in the real system the $d_{z^2}$ orbital may also be slightly hole doped. To be more precise and to enable the doping of the $d_{z^2}$ orbital, we propose a bilayer type II t-J model to describe the low energy physics. The model is a generalization of a model proposed one of us before\cite{zhang2020type,zhang2021fractional}. Basically we take the large $J_H$ limit and restrict to a Hilbert space with four spin 1/2 singlon ($d^7$) states and three spin-one doublon ($d^8$) states.  Inter-orbital $J_H$ disappears in the model with the cost of non-trivial constraint.  The type II t-J model can be understood to describe the low energy physics of doping a spin-one Mott insulator\cite{zhang2022pair} with doped hole in a spin 1/2 state. The model has two important parameters: the total hole doping level $x$ and energy splitting $\Delta$ between the two orbitals to tune the relative doping of the two orbitals.  In the large $\Delta$ limit, we have $n_2=1$ and $d_{z^2}$ is Mott localized and forms a rung singlet. We propose a parton mean field theory to deal with the type II t-J model. In the simple large $\Delta$ limit, in the mean field level we reach a bilayer one-orbital t-J model for an emergent `$d_{x^2-y^2}$' orbital in the mean-field level. In this model, we can automatically get a large  $J_\perp/t$ from our parton mean field theory, justifying our previous phenomenological treatment. From a direct mean field calculation of the type II t-J model, we find s-wave inter-layer pairing at $x=0.5$ similar to the one-orbital t-J model before. 

\textit{Bilayer two-orbital model}: We  start from a two-orbital t-J model on a bilayer square lattice, Fig.~\ref{fig:1} (a), which has the following Hamiltonian,
\begin{eqnarray}
    H&=& H_{K}+ J^x_\parallel \sum_l \sum_{\langle ij \rangle} \vec S_{i;l;1}\cdot \vec S_{i;l;1}+J^z_\perp \sum_{i} \vec S_{i;t;2}\cdot \vec S_{i;b;2} \notag \\
&+ &U'\!\sum_i n_{i;1}n_{i;2}-2J_H\!\sum_i (\textcolor{black}{\vec S_{i;l;1}\cdot \vec S_{i;l;2}}+\frac{1}{4}n_{i;1}n_{i;2}), \label{e1} 
\end{eqnarray}
and
\begin{eqnarray}
    H_{K}&=&-t^x_\parallel \sum_{l ,\sigma} \sum_{\langle i,j \rangle} (P d^\dagger_{i;l;1;\sigma}d_{j;l;1;\sigma} P+H.c.)\notag \\ 
    &-&t^z_\parallel\sum_{l,\sigma }  \sum_{\langle i,j \rangle} (P d^\dagger_{i;l;2;\sigma}d_{j;l;2;\sigma} P+H.c.) \notag \\ 
    &-&t^{xz}_\parallel \sum_{l,\sigma }\sum_{\langle ij \rangle} ((-1)^{s_{ij}} P d^\dagger_{i;l;1;\sigma }d_{j;l;2;\sigma } P+H.c.) \notag \\ 
    &-&t_\perp^z \sum_i (P d^\dagger_{i;t;2;\sigma}d_{i;b;2;\sigma} P+H.c.) +\Delta \!\sum_i (n_{i;1}-n_{i;2}) \notag ,  
\end{eqnarray}
where $P$ is the projection operator to remove the double occupancy of each orbital. Here, $l=t,b$ labels the layer index, and $\sigma=\uparrow,\downarrow$ is for the spin index. We dub $d_1, d_2$ for the $d_{x^2-y^2}$ and $d_{z^2}$ orbital respectively. The hopping parameters are estimated $t_{\parallel}^{x}=0.485$, $t_{\parallel}^{z}=0.110$, $t_{\parallel}^{xz}=0.239$, 
$t_{\perp}^{z}=0.635$ by DFT \cite{luo2023bilayer}.  
$s_{ij}=1$ for the $x$ bond and $s_{ij}=-1$ for the $y$ bond.  For simplicity, we only keep intra-layer $J^{x}_{\parallel}$ for the $d_{x^2-y^2}$ orbital and the inter-layer $J^{z}_{\perp}$ for the $d_{z^2}$ coupling. 
$U'$ is inter-orbital repulsion and $J_H$ is the Hund's coupling.  $n_{i;a}$ is the density for orbital $a=1,2$. \textcolor{black}{$\vec S_{i;l;a}$ is the spin operator for layer $l=t,b$ and orbital $a=1,2$.}  We also ignore the $n_i n_j$ term in the $J$ coupling.
In Fig.~\ref{fig:1}, we illustrate the system and the model. On average we have $n=2-x$ number of electrons (summed over spin) per site with $x\approx 0.5$ in the experiment. We have $n_1 \approx 0.5$ and $n_2 \approx 1$.

\begin{figure}[t]
    \centering
   \includegraphics[width=1\linewidth]{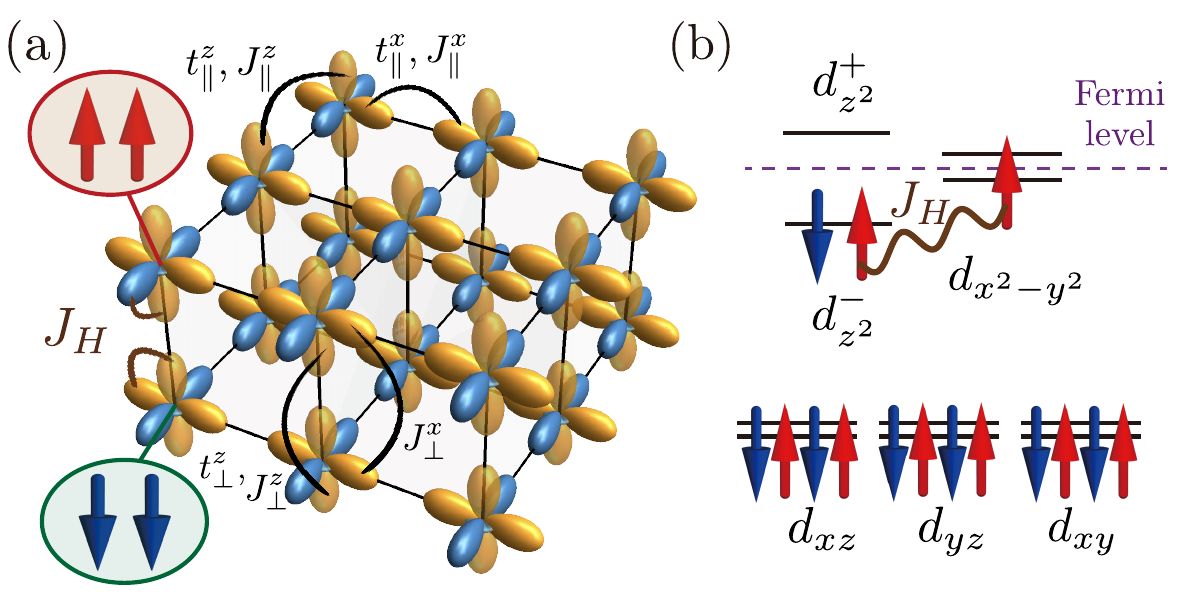}
    \caption{\textbf{(a) The schematics of the bilayer two-orbital model.}
    The various $t,J$'s are introduced for the hoppings and interactions of two orbitals on square lattices.
    Importantly, a strong ferromagnetic Hund coupling $J_{H}$ transmits $J^{z}_{\perp}$ of the $d_{z^2}$ orbital to the $d_{x^2-y^2}$ orbital, by enforcing a spin-triplet at each site (Inset). 
    \textbf{(b) The electronic configuration of two Ni$^{+2.5}$ states in one unit cell.}
    The density per site with summing over spin is roughly  $n_{1}\simeq 1/2$ and $n_{2}\simeq 1$.  
    }
    \label{fig:1}
\end{figure}

\textit{Bilayer one-orbital t-J model}: 
We first consider the limit where the $d_{2}$ orbital is Mott localized with pinned $n_2=1$.  In this limit, $d_{2}$ orbitals form a rung-singlet insulator due to large $J_\perp$ and may be integrated out and one can focus on an one-orbital t-J model with the $d_{1}$ orbital. However, we emphasize that the gapped $d_{2}$ degree of freedom still plays an important role due to the Hund's coupling. A large Hund's coupling enforces the two orbitals to form a spin-triplet at each site. Within the restricted Hilbert space, the spins of the two orbitals align and the inter-layer spin-spin coupling $J_\perp^z$ also induces anti-ferromagnetic coupling of the $d_{1}$ orbital (see the Inset of Fig.~\ref{fig:1}(a)). Basically only the orbital symmetric part, $J^{x}_{\perp}=J^{z}_{\perp}$, can persist in the restricted Hilbert space. 
Consequently, we should consider a significant inter-layer $J_\perp$ also for the $d_{x^2-y^2}$ orbital, though there is no inter-layer hopping.

Motivated by the above considerations, we now consider an effective one-orbital t-J model for the $d_{x^2-y^2}$ orbital, 
\begin{align}
    H_{\mathrm{eff}} &=
-t_{\parallel}^{x} \sum_{l, \sigma}\sum_{ \langle i,j \rangle}
P\left(
d_{i;1;l,\sigma }^{\dagger}\textcolor{black}{d_{1;1;l;\sigma }}
\right) P+H.c. \notag\\
& +J_{\parallel}^{x}\sum_{l}\sum_{ \langle  i, j \rangle}\vec{S}_{i;l;1}
\cdot \textcolor{black}{\vec{S}_{j;l;1}}
+  J_{\perp}^{z}\sum_{i}
\vec{S}_{i;t;1} \cdot \vec{S}_{i;b;1}
\label{eq:one-orbital t-Jmodel}
\end{align}
Hereafter, shorthand notation $t=t^{x}_{\parallel},J_{\parallel}=J^{x}_{\parallel}$, and $J_{\perp}=J^{z}_{\perp}$ are used, unless otherwise stated. Note that the model above is quite unconventional in the sense that we have a large $J_\perp$ but no inter-layer hopping $t_\perp$, compared to other existing models \cite{PhysRevB.95.214509}. This is impossible in the standard t-J model usually with $J<t$. We note a similar model (dubbed as mixed dimensional t-J model) has been proposed in the cold atom context but only out of equilibrium\cite{bohrdt2022strong,hirthe2023magnetically}. 

We then employ the standard U(1) slave-boson mean-field theory\cite{lee2006doping} and represent the electronic operator as, $d_{i;l;1;\sigma}^{\dagger} = f_{i;l;\sigma}^{\dagger} b_{i;l}$ with the constraint $n_{i;l;f}+n_{i;l;b}=1$ (see the Supplemental Material (SM) for details). In mean-field level, we decouple the following order parameters from the $J$ terms: the hopping terms $\chi_{\parallel;ij,\sigma}^{l} = 2 \langle f_{i;l;\sigma}^{\dagger}f_{j;l;\sigma}\rangle$, $\chi_{\perp; i;\sigma}= 2\langle f_{i;t;\sigma}^{\dagger}f_{i;b;\sigma}\rangle$ and  the pairing terms $\Delta_{\parallel;ij}^{l} =  
2s^{ij}\langle f_{i;l;\uparrow}f_{j;l;\downarrow}\rangle$, $\Delta_{\perp;i} =2 \langle f_{i;t;\uparrow}f_{i;b,\downarrow}\rangle$.
We obtain these order parameters from self-consistent calculations. We fix $t_{\parallel}=1$ and $J_{\parallel}=1/2$ and vary the $J_\perp$ and the doping $x$ in the range $0\leq x \leq 1/2$.

Here we summarize our numerical results. In the limit of small $J_\perp$, the model reproduces the well-known behaviors of the single-layer t-J model, with the famous $d_{x^2-y^2}$ pairing within each layer. As the strength of $J_{\perp}$ is gradually increased, there is a first-order transition after which we find s-wave pairing with dominated inter-layer pairing, as illustrated in \textcolor{black}{Fig.\ref{fig:2} (a-b)}. 
\textcolor{black}{In Fig.\ref{fig:2} (c), we find a first-order transition from the d-wave to s-wave pairing with dominated inter-layer pairing.}
With a large enough $J_{\perp}$ (for example, $J_\perp/t$>0.5), the value of $|\Delta_{\perp}|$ remains survives to the large hole doping regime with $x\simeq 0.5$. 

We note that the normal Fermi surfaces are completely gapped in the  s-wave pairing phase, while there are nodes in the d-wave pairing, as depicted in Fig.\ref{fig:2} (d). $J_\perp/t>0.5$ is quite reasonable given that $J_\perp$ origins from the super-exchange of the $d_{2}$ orbital which has a large inter-layer coupling. Thus we expect an s-wave inter-layer paired superconductor in the experimental regime even with a $50\%$ hole doping. We emphasize that it is important to have large $J_\perp$ but with the inter-layer hopping $t_\perp=0$. For example, one can imagine a conventional bilayer t-J model for the $d_{z^2}$ orbital with $t_\perp>J_\perp$. In Fig.\textcolor{red}{S1}  in SM, we show that a large $t_\perp$ term suppresses the pairing because the hopping disfavors inter-layer spin-singlet Cooper pair. Therefore the unusual model we consider here for the $d_{x^2-y^2}$ orbital host has stronger pairing than the usual t-J model.

\begin{figure}[tb]
    \centering
    \includegraphics[width=1\linewidth]{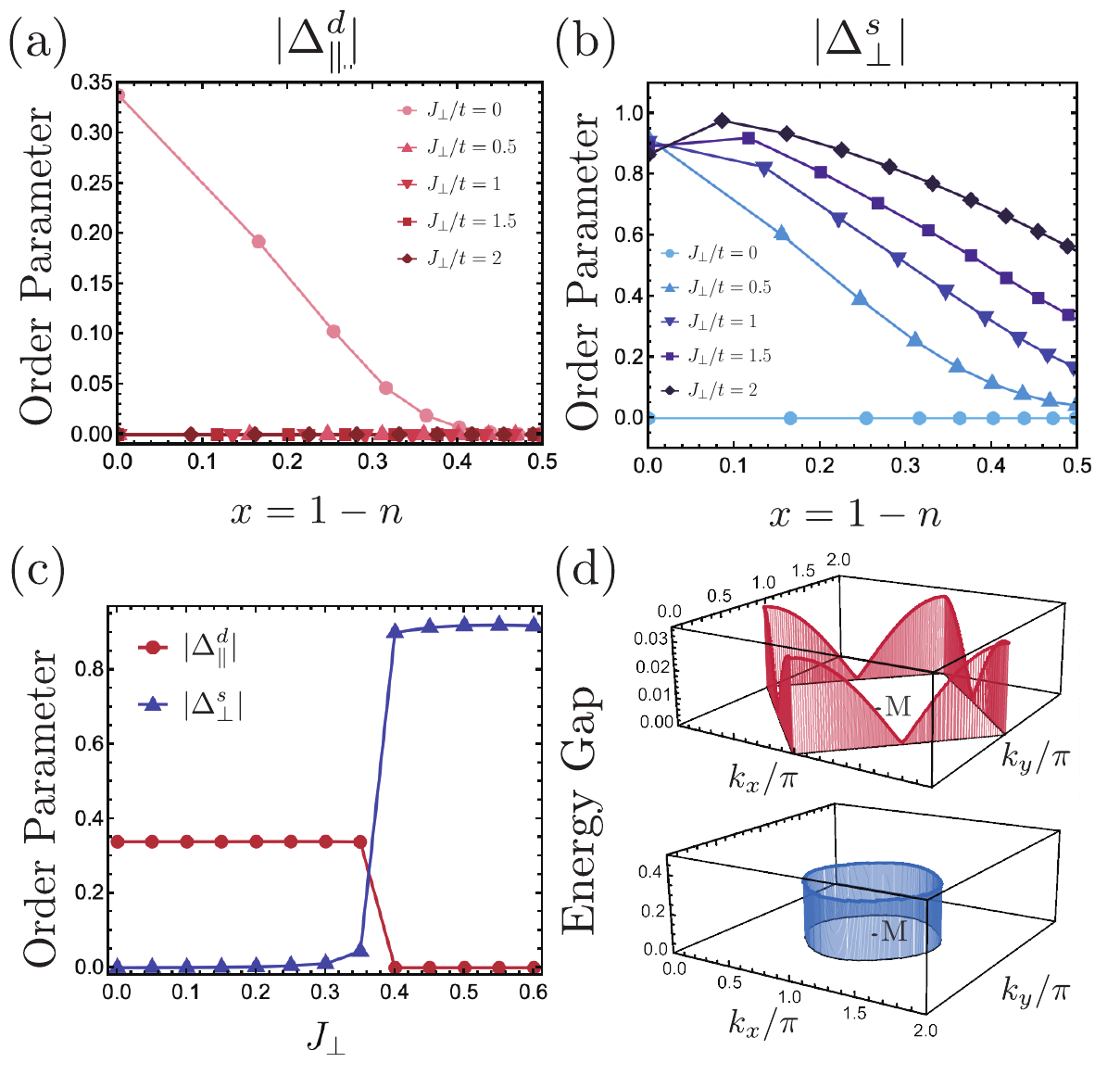}
    \caption{
    \textcolor{black}{
    \textbf{(a-b) Zero temperature mean-field solutions of one-orbital t-J model.} 
 We plot the filling $x$ dependence of (a) intra-layer d-wave pairing, (b) inter-layer s-wave pairing within the slave-boson framework are shown at $t_{\parallel}^{x}=1$, $J_{\parallel}^{x}=1/2$. 
    \textbf{(c) $J_{\perp}$ dependence of pairing order parameter at $x=0$.} 
The inclusion of $J_{\perp}^{z}$ induces the first-order phase transition from $d$-wave pairing, $\Delta_{\parallel}^{d}$, to $s$-wave pairing, $\Delta_{\perp}^{s}$. 
\textbf{(d) The energy gap of the two distinct superconducting states at the Fermi surface.}
Two specific cases of $J_{\perp}^{z}/t^{x}_{\parallel}=0,x=0$ (top) and $J_{\perp}^{z}/t^{x}_{\parallel}=2,x=1/2$ (bottom) are chosen for a illustration.
The normal Fermi surface, centered at the M=($\pi,\pi$) point, is completely gapped with a $s$-wave pairing (bottom), while there are four point nodes with a $d$-wave pairing (top). 
}
}
\label{fig:2}
\end{figure}

\begin{figure}[tb]
     \centering
    \includegraphics[scale=0.5]{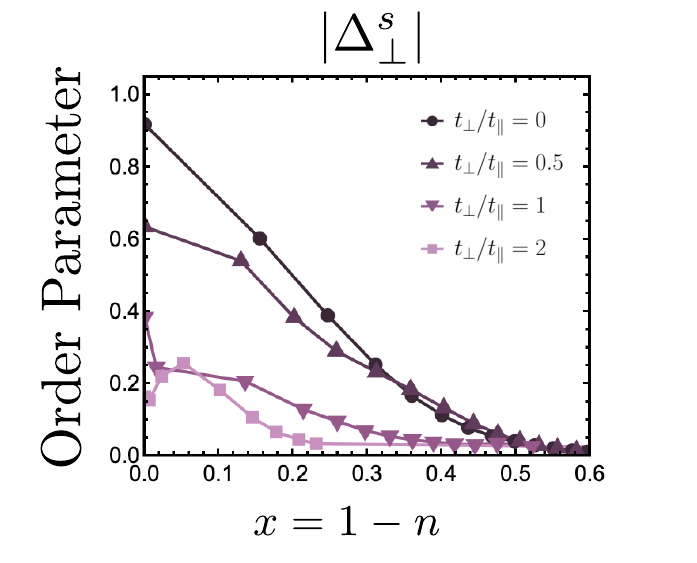}
          \caption{\textbf{
     Mean-field order parameters of the one-orbital model.} Inter-layer hopping $t_{\perp}$ dependence of
the inter-layer pairing at $J_{\perp}=1/2$. The inclusion of larger inter-layer hopping $t_{\perp}$ suppressed the inter-layer pairing order parameter $\Delta_{\perp}$.
     } 

     \label{fig:S1}
 \end{figure}

\textit{Type II t-J model}: The importance of Hund's coupling in sharing the super-exchange $J$ has been demonstrated in the simple case of $n_2=1$ per site.
In this limit, the $d_{2}$ orbital is orbital-selective Mott localized and forms rung-singlet.
Then we just ignore $d_2$ and deal with a one-orbital model and take the transmission of $J_\perp$ by hand.  
However, this approach is not very rigorous and needs a justification. Moreover, in real system, the $d_{2}$ orbital is likely to be slightly hole doped with $n_2<1$. Then the $d_2$ orbital should be kept in the low energy model. In this case, we need to deal with the full two-orbital model in Eq.~\ref{e1}. However, $U'$ and $J_H$ are large and cannot be treated in perturbation or mean field level. Especially, there is no good way to capture the effect of sharing the J terms between the two orbitals from the Hund's coupling. Apparently, a new model and a new method is called for to describe the realistic regimes with two active orbitals and a strong Hund's coupling.

To address this challenging problem, we take a non-perturbative approach. We first take $U',J_H$ to be large and project to a restricted Hilbert space. This leads to a generalization of the type II t-J model proposed by one of us in Ref.\onlinecite{zhang2020type}. 
We only keep four singlon ($d^7$) states and three spin-triplet doublon ($d^8$) states. First, at each site $i$, the four singlon states can be labeled as, $\ket{a \sigma}=d^\dagger_{a;\sigma}\ket{G}$ where $\ket{G}$ is defined as a vacuum states where all $t_{2g}$ orbitals are fully filled with $a=1,2$ and $\sigma=\uparrow, \downarrow$.
Meanwhile, the three spin-triplet doublon states are written as, $\ket{-1}=d^\dagger_{1\downarrow}d^\dagger_{2\downarrow}\ket{G}$, $\ket{0}=\frac{1}{\sqrt{2}}(d^\dagger_{1\uparrow}d^\dagger_{2\downarrow}+d^\dagger_{1\downarrow}d^\dagger_{2\uparrow})\ket{G}$ and $\ket{1}=d^\dagger_{1\uparrow}d^\dagger_{2\uparrow}\ket{G}$. Here, we ignore the site index $i$ for simplicity. The spin-singlet doubly occupied states is penalized by a large $J_H$ and is removed from the Hilbert space.

Now, we project the electron operator inside this $4+3=7$ dimensional Hilbert space, 
\begin{eqnarray}
    d_{i;l;1\uparrow}&= \prod_{j<i}(-1)^{n_j}  \big(\ket{2\uparrow}_{il}\bra{1}_{i\l}+\frac{1}{\sqrt{2}}\ket{2\downarrow}_{il}\bra{0}_{il}\big),
    \label{eq:p_electron_1}
\notag \\
    d_{i;l;1\downarrow}&= \prod_{j<i}\!(-1)^{n_j}  \big(\ket{2\downarrow}_{il}\bra{-1}_{il}\!+\frac{1}{\sqrt{2}}\ket{2\uparrow}_{il}\bra{0}_{il}\big),
     \label{eq:p_electron_2}
\notag \\
    d_{i;l;2\uparrow}&= -\!\prod_{j<i}\!(-1)^{n_j}  \big(\ket{1\uparrow}_{il}\bra{1}_{i\;l}+\!\frac{1}{\sqrt{2}}\ket{1\downarrow}_{il}\bra{0}_{il}\big),
     \label{eq:p_electron_3}
\notag \\
        d_{i;l;2\downarrow}&= -\!\prod_{j<i}(-1)^{n_j}  \big(\ket{1\downarrow}_{il}\bra{-1}_{il}\!+\!\frac{1}{\sqrt{2}}\ket{1\uparrow}_{il}\bra{0}_{il}\big)
     \label{eq:p_electron_4}
\end{eqnarray}
where $\prod_{j<i}(-1)^{n_j}$ is the Jordan-Wigner string. 
The spin operators for the \textit{spin-1/2} singlon state are $ \vec s_{i;a}=\frac{1}{2}\sum_{\sigma \sigma'} \ket{a\sigma}_i \vec \sigma_{\sigma \sigma'}\bra{a\sigma'}_i$ with $\vec \sigma$ as the Pauli matrices. the spin operators for the \textit{spin-one} doublon states are written as $ \vec S_i=\sum_{\alpha,\beta=-1,0,1} \vec T_{\alpha \beta]}\ket{\alpha}_i \bra{\beta}_i$. Here we have $  $, $  T_x=\frac{1}{\sqrt{2}}\begin{pmatrix} 0 & 1 & 0 \\ 1 & 0 & 1 \\ 0 & 1 & 0 \end{pmatrix}$ and $ T_y=\frac{1}{\sqrt{2}}\begin{pmatrix} 0 & -i & 0 \\ i & 0 & -i \\ 0 & i & 0 \end{pmatrix}$ in the $\ket{1},\ket{0},\ket{-1}$ basis.

The type II t-J model Hamiltonian is 
\begin{eqnarray}
     H=H_{K}&+&   J^x_\parallel \sum_l \sum_{\langle ij \rangle} \vec s_{i;l;1}\cdot \vec s_{j;l;1}+J^z_\perp \sum_{i} \vec s_{i;t;2}\cdot \vec s_{i;b;2} \notag \\
&+& J_{sd}^\parallel \sum_l \sum_{\langle ij \rangle} (\vec s_{i;l;1}\cdot \vec S_{j;l} +\cdot \vec S_{i;l}\cdot \vec s_{j;l;1}) \notag \\ 
&+&J_{sd}^\perp 
 \sum_i (\vec s_{i;t;2}\cdot \vec S_{i;b}+\vec S_{i;t}\cdot \vec s_{i;b;2})\notag \\
&+&J_{dd}^\parallel \sum_l \sum_{\langle ij \rangle}\vec S_{i;l}\cdot \vec S_{j;l}+J_{dd}^\perp \sum_i \vec S_{i;t}\cdot \vec S_{i;b} ,
\label{eq:type_II_t_J_main}
\end{eqnarray}
where $H_K$ is the same as in Eq.~\ref{e1}, except that the above projected electron operators are in the $4+3=7$ Hilbert space as defined above. We have $J_{sd}^\parallel=\frac{1}{2}J^x_\parallel$, $J_{sd}^\perp=\frac{1}{2} J^z_\perp$. $J_{dd}^\parallel=\frac{1}{4} J^x_\parallel$ and $J_{dd}^\perp=\frac{1}{4}J^z_\perp$. We are interested in the filling of $n_T=n_1+n_2=1+n=2-x$. If the number of sites is $N_S$, there are $(1-x) N_s$ number of doublon states and $x N_s$ number of singlon states. The energy splitting $\Delta$ in $H_K$ tunes the relative density of the two orbitals. In particular, if $\Delta$ is large and positive, we only need to keep two singlon states corresponding to the $d_2$ orbital.

\textit{Parton mean-field theory}: We employ the three-fermion parton construction\cite{zhang2020type} to deal with the type II t-J model. 
The four singlon states are constructed as  $\ket{a\sigma}_i=f^\dagger_{i;a\sigma}\ket{0}$, while the three S=1 doublons are created by $\ket{-1}_i=\psi^\dagger_{i;1\downarrow}\psi^\dagger_{i;2\downarrow}\ket{0}$, $\ket{0}_i=\frac{1}{\sqrt{2}}(\psi^\dagger_{i;1\uparrow}\psi^\dagger_{i;2\downarrow}-\psi^\dagger_{i;2\uparrow}\psi^\dagger_{i;1\downarrow})\ket{0}$ and $\ket{1}=\psi^\dagger_{i;1\uparrow}\psi^\dagger_{i;2\uparrow}\ket{0}$. We need to impose a local constraint at each site $i$: $n_{i;f}+n_{i;\psi_1}=1$, $n_{i;\psi_1}=n_{i;\psi_2}$ with $n_{i;f}=\sum_{a\sigma}f^\dagger_{i;a\sigma}f_{i;a\sigma}$ and $n_{i;\psi_a}=\sum_\sigma \psi^\dagger_{i;a\sigma}\psi_{i;a\sigma}$. 
On average, we have $n_f=x$ and $n_{\psi_1}=n_{\psi_2}=1-x$ with the convention $n_1+n_2=2-x$. We introduce the notation $\Psi_{i;\sigma}=(\psi_{i;1\sigma},\psi_{i;2\sigma})^T$, then there is another constraint: $\Psi^\dagger_i \vec \tau \Psi_i=0$ where $\vec \tau$ is Pauli matrix in the color space. This constraint enforces the two colors $a=1,2$ forms singlet, thus the spin is in a triplet due to fermion statistics\cite{zhang2020type}. This constraint gives a SU(2) gauge symmetry: $\Psi_i \rightarrow U_i \psi_i$ where $U_i \in SU(2)$ acting in the color space, rotating $\psi_1$ to $\psi_2$.

Within the parton construction, the projected electron operator is represented as,
$d_{i;a\sigma}=\epsilon_{ab}f^\dagger_{i;b \sigma}\psi_{i;2 \sigma}\psi_{i;1\sigma}+\frac{1}{2} \epsilon_{ab} f^\dagger_{i;b\bar \sigma}(\psi_{i;2\downarrow}\psi_{i;1\uparrow}+\psi_{i;2\uparrow}\psi_{i;1\downarrow})
$. Here, $\epsilon_{ab}$ is the anti-symmetric tensor with $\epsilon_{12}=1$ and $\bar \sigma$ denotes the opposite spin of $\sigma$.
The singlon and doublon spin operators are now represented as, $\vec s_{i;a}=\frac{1}{2}\sum_{\sigma,\sigma'} f^\dagger_{i;a\sigma} \vec \sigma_{\sigma \sigma'} f_{i;a\sigma'}$ and $\vec S_i=\frac{1}{2} \sum_{a}\sum_{\sigma \sigma'} \psi^\dagger_{i;a\sigma}\vec \sigma_{\sigma \sigma'} \psi_{i;a\sigma'}$.

Substituting all the above expressions, one can decouple the type II t-J model in Eq.~\ref{eq:type_II_t_J_main} and perform the self-consistent mean-field calculation. We provide all details in SM. In principle, one can have a phase diagram from tuning $\Delta$ and $x$. For simplicity, we her consider the large positive $\Delta$ limit, so that $n_2$ is pinned to be $1$, safely 
ignoring $f_1$ and keeping only the two singlon states occupied by $f_{2\sigma}$. 
This corresponds to orbital selective Mott localization of the $d_{z^2}$ orbital and now $d_{i;2\sigma}=0$ without the $f_{1}$ operator.
One important mean field decoupling is an on-site term, $\langle \psi^\dagger_{i;l;a\sigma}f_{i;l;2\sigma} \rangle =\frac{3}{4}\Phi_a$ for each spin $\sigma$ component. Due to the SU(2) gauge symmetry, we can always fix the gauge to choose $\Phi_2\neq 0$ while $\Phi_1=0$. Then $\langle \psi^\dagger_{i;l;2\sigma} f_{i;l;2\sigma} \rangle =3\Phi_2/4 \neq 0$ and we have $d_{i;l;1\sigma}\sim \frac{3}{4} \Phi_2^\dagger \psi_{i;l;1\sigma}$.  
Now $\psi_{i;l;1\sigma}$ can be identified as the electron operator of the $d_{x^2-y^2}$ orbital with density $n_{\psi_1}=1-x$, while $f_2$ and $\psi_2$ hybridize and form the same band with the total density $n_{f_2}+n_{\psi_2}=1$ per site. They just represent the localized spin moments of the $d_{z^2}$ orbital and form a rung singlet in the bilayer model due to the large $J^z_\perp$ term.

In terms of the emergent `$d_{x^2-y^2}$' orbital $\psi_1$, an effective model can be derived from Eq.~\ref{eq:type_II_t_J_main} by substituting $d_{i;l_1\sigma}\sim \frac{3}{4} \Phi^\dagger_2 \psi_{i;l;1\sigma}$,
\begin{eqnarray}
      H_{\psi_1}&= &\sum_l \sum_{\langle ij \rangle} \Big[
      -\frac{9}{16} |\Phi_2|^2  t^x_\parallel \psi^\dagger_{i;l; 1\sigma} \psi_{i; l; 1\sigma}
      \nonumber
      \\
      &+&
      J^\parallel_{dd} \vec S_{i;l;\psi_1}\cdot \vec S_{j;l;\psi_1}
\Big]
      +J^\perp_{dd} \sum_i \vec S_{i;t;\psi_1}\cdot \vec S_{i;b;\psi_1}  
\end{eqnarray}
where $\vec S_{i;l;\psi_1}=\frac{1}{2} \psi^\dagger_{i;l;1\sigma} \vec{\sigma}_{\sigma \sigma'} \psi_{i;l;1\sigma'}$ is the spin operator of $\psi_1$.  The effective spin-spin coupling for this emergent $\psi_1$ orbital originates from the $J_{dd}$ coupling of the spin-one moments. As a result, the super-exchange of both $d_{z^2}$ and $d_{x^2-y^2}$ orbitals contribute to the $J$ coupling of this effective model. We have a large $J^\perp_{dd}=\frac{1}{4} J^z_\perp$ and large  $J^\parallel_{dd}=\frac{1}{4}J^x_\parallel$ for this emergent $\psi_1\sim d_1$ orbital, even though there is no inter-layer hopping. We also note an interesting effect of reducing the hopping by a factor of $|\Phi_2|^2$ ($|\Phi_2|<0.5$ from our calculation as in Fig \textcolor{red}{S2}(c) in SM).

\begin{figure}[tb]
    \centering
    \includegraphics[width=1\linewidth]{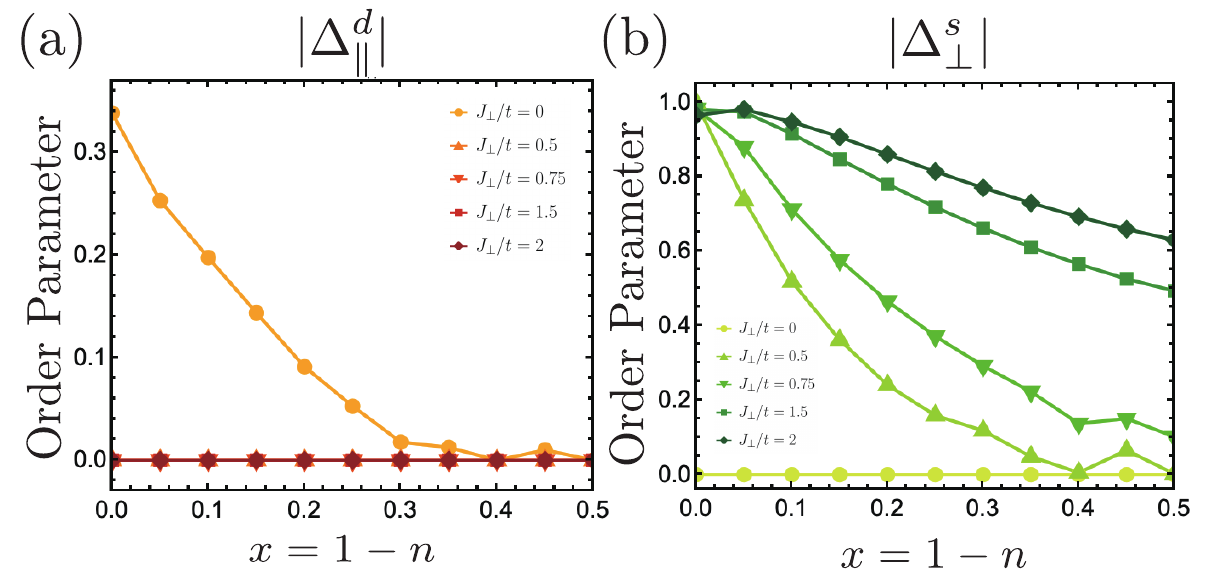}
    \caption{\textcolor{black}{
    \textbf{(a-b) Zero temperature mean-field solutions of type II t-J model in the large $\Delta$ limit.} 
    We plot the filling $x$ dependence of (a) intra-layer pairing, (b) inter-layer pairing of  the emergent `$d_{x^2-y^2}$' orbital at $t_{\parallel}^{x}=1$, $J_{\parallel}^{x}=1/2$. 
Comparing \ref{fig:2}(a-b) and \ref{fig:3}(a-b), we notice that the one-orbital t-J model shows similar behaviors as the more rigorous type II t-J model in the large $\Delta$ limit with the $d_{z^2}$ Mott localized.}
}
\label{fig:3}
\end{figure}

 We perform a full self-consistent mean field calculation involving all $f_2, \psi_1, \psi_2$ orbitals. We confirm that $f_2, \psi_2$ just form a band insulator in agreement with a rung-singlet phase, while the $\psi_1$ orbital is at density $n_1=1-x$ and  gets intra-layer and inter-layer pairing terms as shown in \textcolor{black}{Fig.~\ref{fig:3}(a-b)}. Note that we still use $t$, $J_\parallel$, and $J_\perp$ as abbreviation of $t^x_\parallel$, $J^x_\parallel$ and $J^z_\perp$, and set $t=1$, $J_\parallel=1/2$. Varying $J_\perp$, we again find a first-order transition from the familiar d-wave to s-wave pairing with dominated inter-layer pairing \textcolor{black}{(See Fig.S2(d))}. If we take a large $J_\perp$ such as  $J_\perp/t=1$, the s-wave pairing is still large at $x=0.5$. Overall, the results are qualitatively the same as the previous bilayer one-orbital t-J model (see Fig.~\ref{fig:2}(a-b)), justifying our previous treatment. However, now we achieve these results from a more precise approach of a microscopic model. The sharing of the super-exchange of one orbital to the other orbital is automatically taken care of in our model and parton framework.

\textit{Discussion}: The calculation in Fig.~\ref{fig:2} is limited to the large $\Delta$ regime with the orbital $d_{z^2}$ in a Mott localized state (forming a rung singlet). In the realistic system, we may have  a smaller $\Delta$ and  the $d_{z^2}$ orbital may likely be slightly doped and also participate in the pairing. This will induce some quantitative effects: (1) $d_{z^2}$ orbital also contributes to superconductivity; (2) The effective hole doping level of the $d_{x^2-y^2}$ can get reduced even though the total hole doping level is fixed; (3) The inter-orbital hopping may further transmit the pairing of one orbital to the other orbital.  We note that a two-orbital t-J model has been proposed and studied for La$_3$Ni$_2$O$_7$ (for example, see Ref.~\onlinecite{luo2023bilayer}), but the previous works all ignore the important effect of sharing the super-exchange $J$ coupling between the two-orbitals by the large Hund's coupling. We have demonstrated that this effect is crucial in the large $\Delta$ limit, so obviously it should not be ignored in the smaller $\Delta$ regime.  With both orbitals active, we also can not derive a one-orbital model simply by integrating the $d_{z^2}$ orbital. In this regime, we believe the type II t-J model we propose here is the minimal model to capture all essential ingredients.  A phase diagram of $(\Delta,x)$ can be obtained by extending our parton mean-field theory with $f_1$ orbital included, which we leave to future work.

 \yhz{We also emphasize the difference between our type II t-J model in Eq.\ref{eq:type_II_t_J_main} and the simplified one-orbital $t-J_\parallel-J_\perp$ model in Eq.\ref{eq:one-orbital t-Jmodel}. We here uncover the one-orbital model simply to demonstrate the essence of our mechanism of inter-layer pairing. However, we emphasize here that Eq.\ref{eq:one-orbital t-Jmodel} is not appropriate for Nickelate at least quantitatively even if the $d_{z^2}$ is Mott localized. Starting from the full model in Eq.\ref{e1}, one can reach Eq.\ref{eq:one-orbital t-Jmodel} by integrating the $d_{z^2}$ orbital in the $J_H\ll J^z_\perp$ limit and get $J_\perp\sim \frac{J_H^2}{J^z_\perp}$. But we believe nickelate is in the $J_H\gg J^z_\perp$ limit because Hund's coupling $J_H$ is part of the Coulomb interaction and should be large. Then the perturbative treatment obviously breaks down and we do not see any controlled way to reach the one-orbital t-J model in Eq.\ref{eq:one-orbital t-Jmodel} from Eq.\ref{e1} in the large $J_H$ regime. In the large $J_H$ limit, the appropriate approach is to take the large $J_H$ expansion instead, which leads to our type II t-J model in Eq.\ref{eq:type_II_t_J_main} in the leading order. In the type II t-J model, the localized spin moment from $d_{z^2}$ orbital becomes also dynamical due to the coupling to the holes in the $d_{x^2-y^2}$ orbital. One possible effect is the polaron formation between the hole and the localized spin moment, as has already been demonstrated in a previous study of a 1D type II t-J model\cite{zhang2022pair}. Such polaron effect is completely ignored in the one-orbital t-J model. We believe the type II t-J model is the minimal model to capture all of the essential physics in the nickelate La$_3$Ni$_2$O$_7$. }

\textit{Conclusion}: In summary, we propose and study a bilayer type II t-J model for the superconducting La$_3$Ni$_2$O$_7$ under high pressure. We emphasize the important role of the Hund's coupling between the $d_{x^2-y^2}$ and the $d_{z^2}$ orbital, which enforces the $d^8$ state to be a spin-triplet.  Due to the Hund's rule, the super-exchange of one-orbital can be shared to the other orbital.  We propose a parton mean field treatment of the type II t-J model. In the limit that the $d_{z^2}$ is Mott localized and forms a rung singlet, we reach a bilayer one-orbital t-J model without inter-layer hopping, but with enhanced  inter-layer anti-ferromagnetic spin-spin coupling $J_\perp$ over intra-layer hopping $t$. Mean field theory then predicts a s-wave inter-layer paired superconductor even at hole doping $50\%$, in agreement with the experiment. In future, one natural extension is to tune the orbital splitting $\Delta$ in our type II t-J model to make the $d_{z^2}$ orbital also slightly hole doped. We  also propose future experiments to  reduce $x$ through electron doping to search for an even higher $T_c$ than 80 K.

\textit{Note added}: When finalizing the manuscript, we become aware of a preprint\cite{lu2023interlayer} which also studied a bilayer one-orbital t-J model with strong inter-layer $J_\perp$, \textcolor{black}{which is the same as Eq.\ref{eq:one-orbital t-Jmodel} of our paper. However, in our opinion, the correct model in the large $J_H$ limit is the type II t-J model in the Eq.\ref{eq:type_II_t_J_main} of our paper. These two models are different even when $d_{z^2}$ is Mott localized, see our recent paper \cite{yang2023strong} for comparisons in numerical simulations of these two models. 
}

\textit{Acknowledgement}: YHZ was supported by the National Science Foundation under Grant No. DMR-2237031. 

%

\appendix
\setcounter{equation}{0}
\setcounter{figure}{0}
\setcounter{table}{0}
\setcounter{page}{1}

\maketitle 
\makeatletter
\renewcommand{\theequation}{S\arabic{equation}}
\renewcommand{\thefigure}{S\arabic{figure}}
\renewcommand{\thetable}{S\arabic{table}}

\onecolumngrid
\section{One-orbital t-J model and slave-boson theory}
We start from the one-orbital Hamiltonian, 
\begin{align}
    H &=
-t_{\parallel}^{x} \sum_{l, \sigma}\sum_{ \langle i,j \rangle}
P\left(
d_{i;1;l,\sigma }^{\dagger}d_{1;l;\sigma }
\right) P+H.c. \notag\\
& +J_{\parallel}^{x}\sum_{l}\sum_{ \langle  i, j \rangle}\vec{S}_{i;l;1}
\cdot \vec{S}_{i;l;1}
+  J_{\perp}^{z}\sum_{i}
\vec{S}_{i;t;1} \cdot \vec{S}_{i;b;1},
\end{align}
and perform the mean field theory employing the slave boson representation, $d^{\dagger}_{i;l,1,\sigma}= f^{\dagger}_{i;l;\sigma}b_{i;l}$.
Assuming $\langle b_i \rangle =\sqrt{x}$, after the mean-field decoupling, the mean-field Hamiltonian is given by,
\begin{eqnarray}
H^{MF}_{SB}&=&
-t_{\parallel}
\sum_{l,\sigma,\langle i,j\rangle } \left( f_{i;l;\sigma}^{\dagger}
f_{j;l;\sigma}
+h.c.
\right)
 -t_{\perp} \sum_{\sigma, i}
\left(
f^{\dagger}_{i;t;\sigma}
f_{i;b;\sigma}
+h.c.
\right) 
\\
 &&
  +D_{\parallel}
\sum_{l, \langle i, j \rangle}
\left( 
s_{ij}(f^{\dagger}_{i;l;1;\uparrow}
f^{\dagger}_{j;l;1;\downarrow}
-f^{\dagger}_{i;l;1;\downarrow}
f^{\dagger}_{j;l;1;\uparrow})
+h.c.
\right) \nonumber \\
&&
  +D_{\perp}
\sum_{i}
\left( 
f^{\dagger}_{i;t;\uparrow}
f^{\dagger}_{i;b;\downarrow}
-f^{\dagger}_{i;t;\downarrow}
f^{\dagger}_{i;b;\uparrow}
+h.c.
\right), \nonumber
\label{eq:S_sb}
\end{eqnarray}
with the coefficients,
\begin{eqnarray*}
    t_{\parallel} &=& x t_{\parallel}^{x}+ \frac{3}{8}J_{\parallel}^{x}\chi_{\parallel}
    ,\quad
    t_{\perp} =\frac{3}{8}
    J_{\perp}^{z} \chi_{\perp},\\
    D^{\parallel}&=&
    \frac{3}{8} 
    J^{x}_{\parallel}
    \Delta_{\parallel}^{d},
    \quad
    D^{\perp} = \frac{3}{8} J^{z}_{\perp}\Delta_{\perp}^{s}.
\end{eqnarray*}
There are 4 mean field order parameters,
   \begin{eqnarray}
        \chi_{\parallel}&= &  \sum_{\sigma} \langle 
  f^{\dagger}_{j;l;\sigma}
 f_{i;l;\sigma}
    \rangle  
,\quad
 \chi_{\perp}
    =
     \sum_{\sigma} \langle 
    f^{\dagger}_{i;t;\sigma}
    f_{i;b;\sigma}
    \rangle
   ,\\
   \Delta_{\parallel} &=&
    \langle 
    s^{ij}(f_{i;l;\uparrow}
   f_{j;l;\downarrow}
    -f_{i;l;\downarrow}
    f _{j;l;\uparrow})
    \rangle 
    ,\quad
   \Delta_{\perp}=
    \langle 
    f_{i;t;\uparrow}
    f_{j;b;\downarrow}
    -f_{i;t;\downarrow}
    f _{j;b;\uparrow}
    \rangle. 
\end{eqnarray}
Moreover, the chemical potential should be fixed for conserving the particle number, $n=\sum_{k,l} \langle f^{\dagger}_{k;l;\sigma}
f_{k;l;\sigma}
\rangle =1-x$.

\section{Type II t-J model and Three-fermion parton theory}
We start from the type II t-J model introduced in Eq.\textcolor{red}{4}. Considering the large $\Delta$ limit, the singlon is formed by only $d_{2}$ orbital, thus the Hilbert space is restricted into
$P_{0} = P- \ket{1,\uparrow}
\bra{1,\uparrow}-
\ket{1,\downarrow}
\bra{1,\downarrow}
$. 
In this Hilbert space, electron operators of $d_2$ orbital itself become zero, thus the kinetic Hamiltonian can be expressed in terms of $d_{1}$ orbital, 
\begin{eqnarray}
       H&=&-t^x_\parallel 
    \sum_{l,\sigma,\langle i,j \rangle} (P_{0}d^\dagger_{i;l;1;\sigma}d_{j;l;1;\sigma} P_{0}+h.c.)\\ 
&&+   J^x_\parallel  \sum_{l,\langle i,j \rangle} \vec s_{i;l;1}\cdot \vec s_{j;l;1}
+J^{dd}_\parallel  \sum_{l, \langle i,j \rangle}\vec S_{i;l}\cdot \vec S_{j;l}
+ J^{sd}_\parallel  \sum_{l,\langle i,j \rangle} (\vec s_{i;l;1}\cdot \vec S_{j;l} +\cdot \vec S_{i;l}\cdot \vec s_{j;l;1}) \notag \\ 
&&+J^z_\perp \sum_{i} \vec s_{i;t;2}\cdot \vec s_{i;b;2} +J^{dd}_\perp \sum_i \vec S_{i;t}\cdot \vec S_{i;b} 
+J^{sd}_\perp 
 \sum_i (\vec s_{i;t;2}\cdot \vec S_{i;b}+\vec S_{i;t}\cdot \vec s_{i;b;2})
\nonumber\end{eqnarray}

Here we use the following three-fermion decomposition,
\begin{eqnarray}
d_{i;l;1;\sigma}^{\dagger}&=&
(\psi_{i;l;1;\sigma}^{\dagger}
\psi_{i;l;2;\sigma}^{\dagger})
f_{i;l;2;\sigma}
+\frac{1}{2} (\psi_{i;l;1\uparrow}^{\dagger}\psi_{i;2;l;\downarrow}^{\dagger}
+\psi_{i;1;l;\downarrow}^{\dagger}\psi_{i;2;l;\uparrow}^{\dagger})
f_{i;l; 2;\bar \sigma}
,\\
d_{j;l;1;\sigma}&=&f^\dagger_{j;l;2;\sigma}(\psi_{j;l;2;\sigma}\psi_{j;l;1;\sigma})+\frac{1}{2} f^\dagger_{j;l; 2;\bar \sigma}(\psi_{j;l;2;\downarrow}\psi_{j;l;1;\uparrow}+\psi_{j;l;2;\uparrow}\psi_{j;l;1;\downarrow}).
\label{eq:three_fermion_parton_appendix}
\end{eqnarray}
Employing the standard decoupling principle, the mean-field Hamiltonian is given by 

\begin{eqnarray}
H^{MF}_{TF}&=&
-t_{f;2}
\sum_{l,\sigma,\langle i,j\rangle } \left( f_{i;l;2;\sigma}^{\dagger}
f_{j;l;2;\sigma}
+h.c.
\right)
-
\sum_{a,c=1,2}
t_{\psi;ac}
\sum_{l,\sigma, \langle i,j \rangle}
\left(\psi_{i;l;a;\sigma}^{\dagger}
\psi_{j;l;c;\sigma}
+h.c.\right)
\\
&& -\sum_{a=1,2}
C_{a}^{0}
\sum_{l,\sigma, i}
\left( 
f^{\dagger}_{i;l;2;\sigma}
\psi_{i;l;a;\sigma}
+\psi^{\dagger}_{i;l;a;\sigma}
f_{i;l;2;\sigma}
+h.c.
\right)\nonumber \\
& & -t_{f}^{\perp} \sum_{\sigma, i}
\left(
f^{\dagger}_{i;t;2;\sigma}
f_{i;b;2;\sigma}
+h.c.
\right)
 -\sum_{a,c=1,2}
t_{\psi;ac}^{\perp}
\sum_{\sigma, i}
\left(
\psi^{\dagger}_{i;t;a;\sigma}
\psi_{i;b;c;\sigma}
+h.c.
\right)\nonumber\\
& & -\sum_{a=1,2}
C_{a}^{\perp} \sum_{\sigma, i}
\left(
f^{\dagger}_{i;t;2;\sigma}
\psi_{i;b;a;\sigma}
+
\psi^{\dagger}_{i;t;a;\sigma}
f_{i;b;2;\sigma}
+h.c.
\right)\nonumber\\
 &&
  +D_{\psi;1}
\sum_{l, \langle i, j \rangle}
\left( s_{ij}(
\psi^{\dagger}_{i;l;1;\uparrow}
\psi^{\dagger}_{j;l;1;\downarrow}
-\psi^{\dagger}_{i;l;1;\downarrow}
\psi^{\dagger}_{j;l;1;\uparrow})
+h.c.
\right)\nonumber \\
&&
  +D_{\psi;1}^{\perp}
\sum_{i}
\left( 
\psi^{\dagger}_{i;t;1;\uparrow}
\psi^{\dagger}_{i;b;1;\downarrow}
-\psi^{\dagger}_{i;t;1;\downarrow}
\psi^{\dagger}_{i;b;1;\uparrow}
+h.c.
\right) \nonumber
\\
&& -\mu_{f} \sum_{l,\sigma, i}
f^{\dagger}_{i;l;a;\sigma}
f_{i;l;a;\sigma}
 -\sum_{a=1,2} \mu_{a} \sum_{l,\sigma, i}
 \psi^{\dagger}_{i;l;a;\sigma}
 \psi_{i;l;a;\sigma},\nonumber\label{eq:S_tf}
\end{eqnarray}
with the coefficients,
\begin{eqnarray*}
    t_{\psi;11}&=&
    t_{\parallel}^{x}
    \left[
    \frac{3}{8}
    \chi_{f}\chi_{\psi;22}
    -\frac{9}{16}
    \Phi_{2}^{0}
    \Phi_{2}^{0}
    \right]
+ \frac{3}{8}J^{dd}_{\parallel}\chi_{\psi;11}, 
    \\
    t_{\psi;22}&=&
     t_{\parallel}^{x}
    \left[
    \frac{3}{8}
    \chi_{f}\chi_{\psi;11}
    \right]
+ \frac{3}{8}J^{dd}_{\parallel}\chi_{\psi;22}, 
\quad
    t_{f;2} =t_{\parallel}^{x}
    \left[
    \frac{3}{8}\left( 
    \chi_{\psi;11}\chi_{\psi;22}
    \right)
    \right]
    ,\quad
    C_{2}^{0} =
 t_{\parallel}^{x}
    \left[
    -\frac{9}{8}\Phi_{2}^{0}\chi_{\psi;11}
    \right], \\
    t^{\perp}_{\psi;11}&= &\frac{3}{8}J^{dd}_{\perp}\chi_{\psi;11},
    \quad 
    t^{\perp}_{\psi;22}=\frac{3}{8}J^{dd}_{\perp}\chi_{\psi;22}
    ,\quad 
t_{f}^{\perp} =
\frac{3}{8}J_{\perp}^{z} \chi_{f}^{\perp}
    ,\quad 
C_{2}^{\perp} =
\frac{3}{8}J_{\perp}^{sd} 
\Phi_{2}^{\perp},
\end{eqnarray*}
and
\begin{eqnarray*}
    D_{\psi;1} 
    &=&  \frac{3}{8}J^{dd}_{\parallel} \Delta_{\psi;1}
    ,\quad
    D_{\psi;1}^{\perp}
  = \frac{3}{8}J^{dd}_{\perp} \Delta_{\psi;1}^{\perp}.
\end{eqnarray*}

\begin{figure}[h]
     \centering
    \includegraphics[scale=0.5]{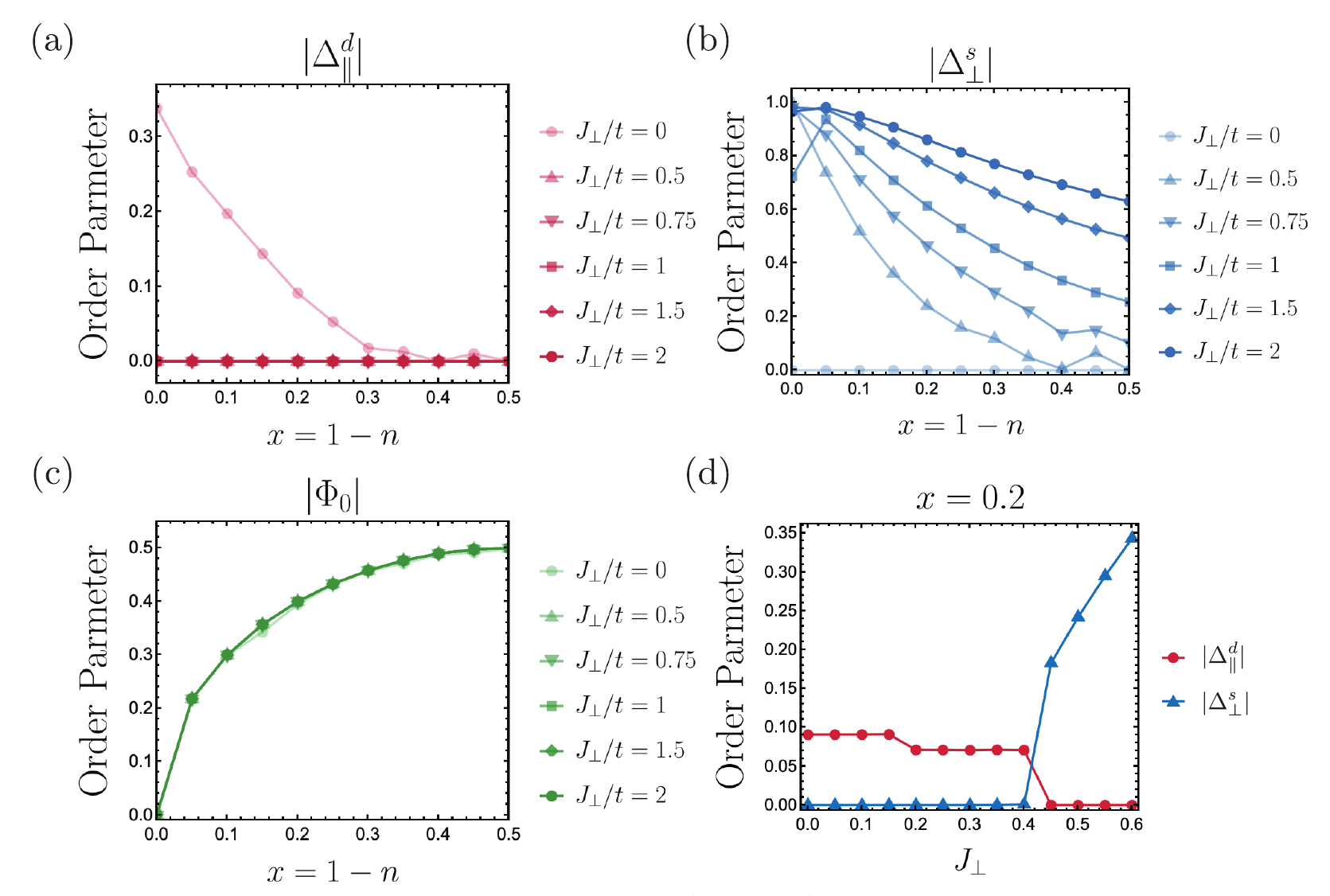}
          \caption{\textbf{
     Mean-field order parameters of the type II t-J model} at $t^{x}_{\parallel}=1$. 
     (a-c) Doping ratio $x$ dependence of intra-layer pairing, inter-layer pairing, Kondo-like coupling at $J_{\parallel}^{x}=1/2$, 
     (d) Inter-layer coupling $J_{\perp}$ dependence of pairings at $x=0.2$.
     }
     \label{fig:S2}
 \end{figure}

\begin{figure}[ht]
    \centering
    \includegraphics[scale=0.55]{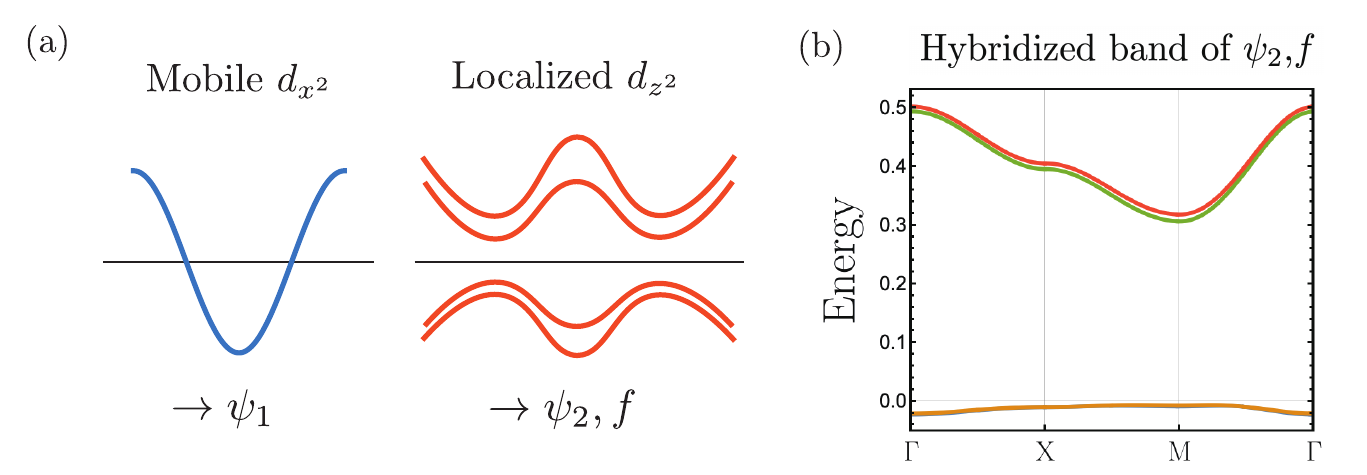}
    \caption{\textbf{(a)Schematic illustrations for physical meaning of three fermions.} $\psi_{1}$ itself means a $d_{1}$ orbital, while $\psi_{2}$,$f$ together form a localized $d_{2}$ orbital.
    \textbf{(b) Energy dispersion of localized $d_{2}$ sector.} We plot the dispersion of the hybridized band of $\psi_{2}$,$f$ for justifying that this sector forms a band insulator in mean field level, indicating a gapped rung-singlet phase. For an illustration, we set $J_{\perp}=1/2,x=0.1$.
    }
    \label{fig:S3}
\end{figure}

There are 10 mean-field order parameters in total for constructing a mean-field Hamiltonian, 
    \begin{eqnarray}
        \chi_{\psi;aa}&= &  \sum_{\sigma} \langle 
    \psi^{\dagger}_{j;l;a;\sigma}
    \psi_{i;l;a;\sigma}
    \rangle  
    ,\quad
    \chi_{f} =\sum_{\sigma} \langle 
    f^{\dagger}_{j;l;2;\sigma}
    f_{i;l;2;\sigma}
    \rangle  
    ,\quad
    \Phi_{2}^{0} =\sum_{\sigma} \langle 
    \psi^{\dagger}_{i;l;2;\sigma}
    f_{i;l;2;\sigma}
    \rangle,
\\
 \chi^{\perp}_{\psi;aa} 
    &=& 
     \sum_{\sigma} \langle 
    \psi^{\dagger}_{i;t;a;\sigma}
    \psi_{i;b;a;\sigma}
    \rangle
    ,\quad 
        \chi_{f}^{\perp}= \sum_{\sigma} \langle 
    f^{\dagger}_{i;t;2;\sigma}
    f_{i;b;2;\sigma}
    \rangle  
,\quad
    \Phi_{2}^{\perp}
    =
    \sum_{\sigma} \langle 
    \psi^{\dagger}_{i;t;2;\sigma}
    f_{i;b;2;\sigma}
    \rangle,\\
   \Delta_{\psi;1} &=&
    \langle 
    s^{ij}(\psi_{i;l;1;\uparrow}
    \psi_{j;l;1;\downarrow}
    -\psi_{i;l;1;\downarrow}
    \psi _{j;l;1;\uparrow})
    \rangle 
    ,\quad
   \Delta_{\psi;1}^{\perp}=
    \langle 
    \psi_{i;t;1;\uparrow}
    \psi_{j;b;1;\downarrow}
    -\psi_{i;t;1;\downarrow}
    \psi _{j;b;1;\uparrow}
    \rangle. 
\end{eqnarray}
Note that $t_{\psi;12}=C^{0}_{1}=C^{\perp}_{1}=\chi_{\psi;12}=\Phi_{1}^{0}=\Phi_{1}^{\perp} =0$, and $J_{sd}^\parallel=\frac{1}{2}J^x_\parallel$, $J_{sd}^\perp=\frac{1}{2} J^z_\perp$, $J_{dd}^\parallel=\frac{1}{4} J^x_\parallel$, $J_{dd}^\perp=\frac{1}{4}J^z_\perp$.  
Together with the order parameters, one should impose the constraints on the number of fermion $n_{\psi;1}=n_{\psi;1}=1-x$, and $n_{f}=x$, where the particle numbers are defined as, 
\begin{eqnarray*}
n_{\psi;a}=\sum_{k,l} \langle \psi^{\dagger}_{k;l;a;\sigma}
\psi_{k;l;a;\sigma}
\rangle ,\quad
n_{f}=\sum_{k,l} \langle f^{\dagger}_{k;l;2;\sigma}
f_{k;l;2;\sigma}\rangle.    
\end{eqnarray*}
In Fig.\ref{fig:S2}, we plot $(\Delta^{\parallel}_{\psi;1}, \Delta^{\perp}_{\psi;1},\Phi^{0}_{2})$ upon doping with a fraction $x$ of holes. 
Moreover in Fig.\ref{fig:S3}, we illustrate the physical meaning of the three fermions in our parton construction. With a non-zero $\Phi=\Phi^0_2$, the  $\psi_{1}$ orbital can be identified as the $d_{1}$ orbital from Eq.~\ref{eq:three_fermion_parton_appendix}. At the same time, $\psi_{2}$,$f$ together form a localized $d_{2}$ orbital with total density $n_{i;2}+n_{i;f}=1$ per site. In our bilayer model they form a gapped rung-singlet phase.

\end{document}